\documentclass[
  english,        
  font=times,     
  twocolumn,      
]{tumarticle}

\usepackage[hyphenbreaks]{breakurl}
\usepackage[T1]{fontenc} \usepackage{ae} \usepackage{aecompl}
\usepackage{lipsum}
\usepackage{balance}
\usepackage[english]{babel}
\addto{\captionsenglish}{\renewcommand{\abstractname}{Executive Summary}}

\title{Europe’s AI Imperative}
\subtitle{A Pragmatic Blueprint for Global Tech Leadership}

\author[affil=1, email=Gjergji.Kasneci@tum.de]{Gjergji Kasneci}
\author[affil=1, email=Urs.Gasser@tum.de]{Urs Gasser}
\author[affil=1, email=Thomas.Hofmann@tum.de]{Thomas F. Hofmann}
\author[affil=1, email=Gerhard.Kramer@tum.de]{Gerhard Kramer}
\author[affil=1, email=Gerhard.Mueller@tum.de]{Gerhard Müller}
\author[affil=1, email=Claudia.Peus@tum.de]{Claudia Peus}
\author[affil=1, email=Helmut.Schoenenberger@tum.de]{Helmut Schönenberger}
\author[affil=1, email=Enkelejda.Kasneci@tum.de]{Enkelejda Kasneci}

\affil[mark=1]{Technical University of Munich}

\date{February 11, 2025}

\begin{document}

\maketitle

\begin{abstract}

Europe is at a make-or-break moment in the global AI race, squeezed between the massive venture capital and tech giants in the US and China’s scale-oriented, top-down drive. At this tipping point, where the convergence of AI with complementary and synergistic technologies, like quantum computing, biotech, VR/AR, 5G/6G, robotics, advanced materials, and high-performance computing, could upend geopolitical balances, Europe needs to rethink its AI-related strategy. On the heels of the AI Action Summit 2025 in Paris, we present a sharp, doable strategy that builds upon Europe’s strengths and closes gaps. 

First, we urge swift scale-up of semiconductor capacity and HPC infrastructure via the EU Chips Act and EuroHPC, giving startups and established firms a launchpad for AI breakthroughs. Second, we propose a European AI Coordination Council, a lean, coordination-focused body to unify standards, enable sector-specific best-practice regulatory frameworks, promote interoperability, and enable agile light-touch regulatory sandboxes for quick prototyping in high-risk AI use cases. Third, we highlight competitive incentive mechanisms, such as tax credits, matching funds, and innovation vouchers instead of heavy-handed direct grants or low-interest loans, to unleash SME innovation in areas where Europe already leads, from advanced manufacturing to green tech.

To retain top talent, our plan coordinates universities and industry for real-world, lifelong training, flexible career pathways, and an innovation-driven R\&D climate. Responsible governance is baked in through practical AI audits, transparent labeling, and credible enforcement of audits for high-risk systems. Finally, we underscore alliances to shape global standards on AI, quantum, robotics, and next-gen connectivity. By fusing these steps into an immediate, well-resourced roadmap, Europe can leap beyond “regulatory first mover” status and become the world’s definitive leader in responsible AI development and usage, powerfully competitive and firmly rooted in Europe's core values.

\end{abstract}

\section{Introduction}
\label{sec:introduction}

Europe is at a “now-or-never” moment in the global AI race, squeezed between the United States’ massive venture-capital ecosystem and China’s scale-oriented, top-down drive. At this crossroads, as \textbf{AI is rapidly converging with complementary and synergistic technologies, such as quantum computing, biotech, VR/AR, 5G/6G, robotics, advanced materials, and high-performance computing, industries and geopolitical balances can be radically reshaped}. Drawing on Europe’s collaboration frameworks (such as Horizon Europe) and established regulatory instruments, such as GDPR, AI Act, DSA, Data Governance Act~\cite{GDPR2016,EUAIACT2021,DSA2020,DataGov2022}, as well as its industrial backbone of “hidden champions”, this article offers a targeted, executable roadmap for Europe's technology lead in AI. We believe now is the time to accelerate as European industries push for swift AI adoption \cite{EUChampionsInitiative} and the AI Action Summit in Paris sparked record investments of €150 billion from private companies plus another €50 billion from the InvestAI Initiative~\cite{AIActionSummit2025,VonDerLeyen}.

We emphasize three urgent pillars for Europe’s competitiveness:

\textbf{Semiconductor Capacity and GPU-Accelerated HPC.} Europe’s share of global semiconductor production is under 10\%, while its GPU resources significantly trail those of the US and Asia \cite{ECChipsAct}. According to recent HPC rankings, Europe’s presence in the TOP500 is commendable but still considerably overshadowed by both the US and China \cite{HPCTOP500, EuroHPCSummit2023, EuroHPC}. Rapidly bridging this gap via new foundry investments, strategic foreign partnerships, and open-source hardware consortia is indispensable for technological sovereignty. We advocate for market-based incentives and competitive frameworks rather than direct state control. 

\textbf{Integrate and Strengthen Existing Frameworks.} Modeled on the successful coordination practices of leading innovation hubs, an innovation-focused European AI Coordination Council would serve as an advisory body to unify standards, tackle fragmentation, and pilot light-touch regulatory sandboxes for quick prototyping. Leveraging the AI Act, it would also promote a voluntary and incentive-driven AI Trust Certificate through third-party certifications, upholding safe and responsible systems. 

\textbf{Competitive, Market-Facilitated Support for SMEs and Mid-Sized Innovators.} Europe’s high-impact growth potential depends on bridging the “Valley of Death” \cite{ValleyOfDeathEU2024} through competitive incentives such as tax credits, matching funds, and innovation vouchers rather than heavy direct subsidies. By linking universities and academic research to SMEs in real-world “living labs”, and leveraging privileged access to sandboxes under the EU AI Act, Europe can convert its academic excellence into global market results.

Talent retention and recruitment are equally decisive in achieving full competitive power in AI and related fields. We propose \textbf{ university-industry programs}, e.g., incentivized through matching funds, flexible R\&D career paths, and targeted immigration reforms to secure a pipeline of world-class talents and AI experts. Meanwhile, \textbf{dynamic but practical governance}, covering sector-specific AI audits, labeling of high-risk systems, and sector-specific compliance sandboxes, ensures that rigorous standards do not stifle Europe’s ability to innovate swiftly.

Finally, Europe must forge \textbf{global alliances} to shape rules and guardrails around \emph{generative AI} (AI systems that create content such as text, images, or audio) and \emph{foundation models} (versatile, large-scale models that can be adapted to multiple downstream tasks), quantum, and next-gen connectivity. By weaving together these elements, ambitious legislation, robust infrastructure, smart regulation, and a unifying vision, Europe can transcend its role as a “regulatory first mover” and become the definitive global catalyst for responsible AI. 

Moreover, by emphasizing Europe’s own strengths, this strategy demonstrates how Europe can differentiate itself from US venture-capital-driven rapid iteration and from China’s top-down scale deployments.

\section{The Global AI Landscape and Europe’s Position}
\subsection{Comparative Dynamics}
The pursuit of AI leadership is at a genuine tipping point: advancements in Generative AI, Quantum Computing, VR/AR, 5G, 6G, and Robotics increasingly amplify one another, raising global stakes. For Europe, this is not an optional race but a \textit{strategic imperative}. The United States and China together pour hundreds of billions into scaling AI research and deployment, exemplified by the \$500 billion Stargate private-joint venture \cite{OpenAIStargate}, fully aware that AI’s economic and geopolitical impact will be pivotal for decades. \textbf{Europe must match this with bold, targeted, and urgent commitments and unwavering resolve}.

\textbf{United States.} Since 2022, the US AI landscape has evolved dramatically. In 2022, federal spending on AI reached \$3.3 billion while private investments hit \$47.4 billion, and by 2023 these investments had surpassed \$67 billion. This surge spurred a vibrant startup ecosystem, raising nearly \$50 billion~\cite{aiindex2023,reuters2023,andre2024ai}. The deployment of the Frontier supercomputer in 2022, e.g., achieving 1.1 exaflops, set the stage for next-generation systems like Aurora and El Capitan that are now revolutionizing high-performance computing for AI research~\cite{HPCTOP500}. Concurrently, landmark regulatory actions, including the NIST AI Risk Management Framework and a comprehensive executive order on AI safety, established robust guidelines for ethical and secure AI development~\cite{nist2023,whitehouse2023}. Together, these initiatives have reinforced the US position as a global leader in AI and set international standards for investment, innovation, and governance.

\textbf{China.} Since 2022, China's AI landscape has experienced rapid expansion fueled by significant state-led investments and a vibrant private sector. Just recently, China unveiled a 1~trillion yuan ($\approx$\$140~billion) financing plan to boost AI development, while private funding in 2023 reached roughly \$7.8~billion~\cite{crunchbase2023_china,crunchbase2023_china_q2}. Major tech giants such as Baidu, Alibaba, Tencent, and Huawei have advanced homegrown AI models, e.g., Baidu’s ERNIE Bot, Alibaba’s Tongyi Qianwen, and Tencent's Hunyuan, which are transforming industries from autonomous driving to e-commerce. Concurrently, China has bolstered its high-performance computing infrastructure with exascale systems such as Sunway OceanLight and Tianhe-3, spurring robust domestic AI chip development in response to US export controls~\cite{top5002024}. Comprehensive regulatory measures, including algorithmic recommendation rules, deep synthesis regulations, and generative AI service controls introduced since 2022, have established a governance framework aligning AI growth with national priorities~\cite{chinaBriefingGenerativeAI}. Together, these initiatives have cemented China's position as a serious global competitor in AI, narrowing the gap with the US while advancing its own innovation ecosystem.

\textbf{Europe.} Europe’s HPC and semiconductor deficits, compounded by scaling challenges and the initial funding bottlenecks for flagship projects and startups like Mistral and AlephAlpha, threaten to undercut its regulatory leadership, e.g., GDPR, AI Act, DSA~\cite{startupverband2024generativeai}. Annual AI-related venture capital investment in the EU is approximately eight times less than in the US and around 50\% of China's investments~\cite{fratto2024scale, ValleyOfDeathEU2024, EUInvestsAI}.

Europe's high-performance computing capacity (such as the LUMI system in Finland and Leonardo in Italy) is indeed considerable~\cite{EuroHPC,HPCTOP500}. Still, it is not sufficient for the requirements of cutting-edge AI training \cite{Martens2024}. This shortfall is partly due to a fragmented infrastructure and a lack of investment in HPC resources over the years \cite{Gigler2018}. However, initiatives are underway to bridge this gap: the EuroHPC Joint Undertaking~\cite{EuroHPCSummit2023}, national exascale projects like the \textit{JUPITER} supercomputer~\cite{dumiak2023exascale}, and new “AI factories” are aimed at significantly boosting Europe's computing capabilities \cite{EuroHPC}. Furthermore, the recent introduction of DeepSeek-V3~\cite{liu2024deepseek}, China’s cost-effective AI model that has upended market expectations by delivering impressive performance at only a fraction of the investment, vividly demonstrates that breakthrough innovations in AI need not rely on enormous budgets. This disruptive development currently reinforces the AI race and offers Europe a renewed opportunity to capitalize on its strengths, thereby facilitating market-driven innovation and private-sector leadership.

\subsection{Europe’s Strengths are Europe’s Dilemmas in the Age of AI}
\textbf{Regulatory Pioneering.} The GDPR showed Europe’s capability to influence global data protection. Together with the AI Act, Europe is well-positioned to lead on responsible AI standards. However, Europe's AI governance model is increasingly being put to the test. Mario Draghi's 2024 report on the economic competitiveness of Europe warned that excessive regulation, unclear enforcement mechanisms, and burdensome compliance cause AI talent and investment to migrate out of the EU; while the AI Act aims to establish trust and accountability, critics argue that unclear enforcement mechanisms and complex compliance processes could deter smaller companies from using AI at scale~\cite{draghi2024a,draghi2024b}.
 
\textbf{Human-centric Leadership \& Societal Commitment.} European institutions spotlight algorithmic transparency, bias mitigation, and privacy \cite{HLEG2019}, a stance increasingly valued worldwide. Transforming discussions into actionable measures necessitates standardized auditing protocols and industry-specific checklists, supported by a European AI
Coordination Council and third-party auditing agencies following sector-specific best-practice frameworks.

Similarly, social responsibility, sustainability, and inclusion are ingrained in European business and governance models. Still, these ideals must manifest in \emph{large-scale pilots, living labs}, and \emph{accessible AI tools} for SMEs to translate high-level objectives into everyday economic gains.

\textbf{Research Leadership.} Europe is a research powerhouse and home to some of the world's leading universities and research institutions, which have been at the forefront of groundbreaking AI research. Innovations like Stable Diffusion~\cite{rombach2022high}, and transformative projects led by HuggingFace~\cite{huggingface2022} and Mistral~\cite{mistral2024} exemplify Europe's extraordinary potential in artificial intelligence. However, despite these high-impact research achievements, the translation of academic breakthroughs into scalable, industrially relevant innovations remains marginal and slow. The gap between research and market application is widened by fragmented funding mechanisms, regulatory hurdles, and a lack of cohesive technology transfer frameworks~\cite{draghi2024a,draghi2024b}. Consequently, while foreign tech giants rapidly commercialize cutting-edge AI innovations, Europe risks underutilizing its academic excellence and losing some of its brightest minds to more dynamic markets abroad. Over time, this disconnect between groundbreaking research and real-world application could significantly erode Europe's competitive edge in AI research.

\textbf{Hidden Champions in the SME Landscape.} Europe’s Mittelstand (99\% of businesses, employing two-thirds of its workforce) often dominates niche global markets (e.g., industrial robotics, agritech). However, AI adoption among SMEs remains low, with barriers including: 
  \begin{itemize}
      \item \emph{Lack of expertise}: Many SMEs lack in-house AI specialists~\cite{thinkbusiness2025} and depend on external AI service providers.
      \item \emph{Cost constraints}: AI implementation requires significant upfront investments, which most SMEs struggle to finance.
      \item \emph{Fragmented access to AI tools}: Unlike in the U.S., where AI-as-a-service models dominate, European AI solutions remain sector-specific and less integrated.
      \item \emph{Lack of AI-related governance and regulatory expertise}: There is an urgent need for agile, forward-thinking governance models that balance regulatory and ethical imperatives, real-time risk management, and cross-border harmonization, enabling SMEs to adopt and scale AI solutions confidently.
  \end{itemize}

To address these issues, the EU has initiated several actions: 
    
\textit{AI Scale-Up Fund:} Grants and low-interest loans to overcome the “Valley of Death” \cite{ValleyOfDeathEU2024, EUAIResearchFund}.
     
\textit{Regulatory Sandboxes:} Sector-specific prototyping environments for high-risk AI use cases (e.g., healthcare, financial services) with accredited auditors overseeing reduced bureaucracy~\cite{madiega2022artificial}. 
     
\textit{Leverage Local Strengths and Regional Diversity:} Identify strategies to encourage SMEs to localize AI solutions (e.g., advanced robotics for mid-sized factories), capitalizing on Europe’s regional plurality~\cite{EESC2021}.
    
However, to fully unlock these possibilities, Europe must secure data and infrastructure sovereignty, underpinned by robust, GPU-accelerated high-performance computing that is readily accessible to SMEs and research institutions, alongside streamlined, forward-thinking compliance frameworks -- all of this underlined with first-class educational and talent retention strategies. The reward: empowered “hidden champions” that could propel Europe’s AI-driven economic transformation in line with Europe's regulations.

\section{Strategic Imperatives for European AI Governance}
Despite world-class research, Europe lags behind the US and China in the commercialization, funding, and fundamental models of AI. While the US is leading with platforms like OpenAI and China is supporting large AI megaprojects, \textbf{Europe's late start in large-scale generative AI requires urgent reform}. To secure its potential for trusted AI governance, Europe must streamline regulation to increase development agility and reflect sector-specific challenges and needs. In addition, Europe must boost investment and considerably expand access to computing power. Below, we propose concrete action points integrating prior regulatory frameworks with urgent on-the-ground needs.

\subsection{Integrate and Strengthen Existing Frameworks}
Though the EU has introduced holistic digital regulations (GDPR, AI Act, DSA, Data Governance Act), coherence, adaptability, and above all, the empowerment of large-scale industrial applications are paramount:

\textbf{Interoperable, Sector-Focused Rules.} We believe it is crucial to have guidelines tailored to verticals such as healthcare, manufacturing, and creative industries, as each of these verticals comes with sector-specific challenges and needs. These vertical guidelines need to work together more efficiently with horizontal regulations. Risk and quality analysis, as well as corresponding measures for AI solutions, vary vastly across sectors and use cases. Therefore, and consistent with the AI Act, it is important to incentivize interoperability, including adoption of standards and industry best practices, by offering co-funded AI projects to Member States. 

\textbf{Targeted Coordination Mechanisms.} There is a need not only for investment in technological innovation, but also innovation in new types of institutions. Modeled after the successful coordination practices of leading innovation hubs, we propose the formation of a European AI Coordination Council (drawing inspiration from DARPA in the US) to serve as an advisory body to derive meaningful interoperability among the broad range of AI norms and standards (in close collaboration with industry experts) and unify them where possible, tackle fragmentation, and pilot light-touch regulatory sandboxes for quick prototyping in high-risk domains. Leveraging the AI Act, this Council would also promote a voluntary and incentive-driven AI Trust Certificate through third-party certifications and drive sector-specific best-practice frameworks for safe and responsible systems.

\textbf{Regulatory Learning} In particular, we recommend recalibrating definitions and risk levels in 18-month regulatory review cycles for rapidly changing areas such as generative AI, AI- and quantum-computing-empowered search algorithms. By integrating structured feedback from industry, academia, and regulatory experts, and systematically synthesizing insights from distributed regulatory sandboxes, such an agile approach could help minimize lag, reduce uncertainty for innovators, and ensure that oversight keeps pace with technological change.

\textbf{Global Capacity-Building.} We propose to engage not only with international bodies such as the OECD, G20, and specialized standard-setting organizations, but also to intensify collaborations with majority world countries. This can be achieved through university partnerships and initiatives facilitated by regional development banks and national development agencies to enhance bilateral knowledge exchange and build robust networks and communities of practice, such as current efforts to establish and connect AI Safety Institutes globally. This approach will help position European values and frameworks as de facto global benchmarks as we move from a technological to a regulatory arms race in the realm of emerging technologies.

\subsection{Achieve Technological Sovereignty through Industrial Policy}
European AI success rests on hardware autonomy and dynamic R\&D:

\textbf{Closing the Semiconductor Gap.} Via the \textit{Chips Act} \cite{ChipsAct2023}, Europe must aim to boost its share of global semiconductor production to at least 20\%, relative to its economic scale, potential, and needs. The focus should be on GPUs and AI accelerators through open-source hardware consortia, new fabrication plants, and strategic alliances (e.g., joint ventures with non-EU foundries).

\textbf{HPC Acceleration.} EuroHPC \cite{EuroHPC} remains crucial and an excellent initial example of Europe's commitment to developing world-class HPC infrastructure. However, Europe must considerably expand its GPU-centric computing capacity. As of 2024, Europe’s representation in the TOP500 list of the world's most powerful supercomputers stands at roughly 28\%, trailing the US and China \cite{HPCTOP500}. For future-proof AI development and scale, it is crucial to encourage exascale and GPU-centric investments and regionally distributed HPC clusters accessible to SMEs and research institutions.

\textbf{Federated Data Platforms.} Building on the practical insights from the Gaia-X Project \cite{GAIA-X}, Europe should incentivize the development of federated data platforms that enforce open standards and user-friendly APIs to enable secure, real-time data sharing. These platforms must integrate HPC frameworks for distributed training on local datasets, ensuring data sovereignty and privacy while providing advanced AI capabilities to SMEs and research institutions. 

\subsection{Competitive, Market-Facilitated Support for SMEs and Mid-Sized Innovators.}
We believe that Europe's high growth potential depends on effectively bridging the “valley of death” through competitive, market-based financial mechanisms rather than high direct subsidies. To achieve this, we recommend the following targeted measures to reduce barriers to early-stage investment and promote breakthrough innovations.

\textbf{Tax credits.} Tax-related incentives can lower the effective costs of innovation by reducing the financial burden on companies investing in advanced AI research and development. These credits reduce risk and incentivize both startups and established companies to allocate resources to breakthrough technologies, thus accelerating progress and market impact.

\textbf{Matching funds.} Such co-investments could play a crucial role by leveraging public investment with private capital. They can ensure that public funds are used efficiently and serve as a catalyst for projects with high market potential, thereby not only boosting the confidence of private investors but also helping to secure the capital needed to bring innovative ideas from research prototypes to market maturity.

\textbf{Innovation vouchers.} This kind of support could serve as a practical policy tool, granting SMEs direct access to specialized expertise and essential resources. By funding short-term or high-impact projects, such vouchers can be used to mitigate regulatory or financial risks of early-stage innovation, enabling firms to test and refine their ideas cost-effectively and fostering sustainable growth.

\textbf{Linking universities and academic research to SMEs.} Fostering academic-industry collaborations through specific funds and real-world living labs would establish a dynamic platform for rapidly translating theoretical findings and research breakthroughs into market-ready applications. A close collaboration between academia and industry not only aligns research with practical needs and competitive market conditions but could also catalyze the commercialization of academic excellence towards a resilient innovation ecosystem in Europe. Furthermore, fostering a venture-friendly ecosystem by streamlining IP transfer rules from universities to industry could spur the timely commercialization of AI breakthroughs. The establishment of strong startup and AI factories in leading university ecosystems around Europe would accelerate the creation of new AI companies. On top, European or national initiatives like WIN in Germany or Tibi in France could help unlock private venture and growth capital for the next generation of AI scale-ups~\cite{startupfactories,kfwWINinitiative,tresorFourthIR}.

\section{Other Crucial Considerations}

\subsection{Education, Reskilling, Talent Retention, and Public Engagement}
The rapid evolution of AI necessitates continuous talent cultivation~\cite{kasneci2023chatgpt} and robust public engagement. To this end, we propose the development of industry-aligned and interdisciplinary curricula through strategic partnerships with organizations such as EIT Digital~\cite{EITDigital} and leading universities, offering intensive AI as well as interdisciplinary degrees and fellowships that provide private-sector sponsors with priority access to emerging talent. Complementary reskilling initiatives (delivered both online and in-person) and upskilling programs should target sectors vulnerable to automation, such as manufacturing and logistics, with small grants available to support workforce retraining. Furthermore, streamlining visa pathways, offering competitive research grants, and establishing flexible “talent bridging” tracks from academia to industry are essential to attracting and retaining top talent.

Simultaneously, public trust in AI must be cultivated through transparent communication and active civic participation. Regular citizen assemblies, including town halls and online forums, should be organized to discuss pressing issues, such as the implications of fake content generation, and incorporate community insights into policymaking. Additionally, partnerships with broadcasters to rigorously label AI-generated content (including deepfakes) will enhance media and AI literacy, while targeted educational campaigns in schools and adult learning programs will ensure that the broader public remains well-informed and engaged with AI developments.

\subsection{Global Alliances and Standard-Setting for Future Technologies}
Europe must leverage its regulatory influence to shape global AI norms and standards. Europe can promote frameworks that embody transparency, accountability, and human-centricity by actively engaging with international bodies such as the OECD, G20, ITU, and specialized standard-setting organizations, and advancing partnerships and initiatives such as the Global Network of AI Safety Institutes and Policy Clinics. Initiatives like establishing a ``G10 for Responsible AI''~\cite{Mazzucato2018} would position European models as global lodestars. In parallel, science diplomacy efforts aimed at bridging the global digital divide, including multistakeholder initiatives such as GESDA~\cite{GESDA}, will help share Europe's balanced vision of AI governance worldwide, ensuring that its standards resonate across borders.

\subsection{Future-Proofing Against Emerging Technology Paradigms}
Europe must proactively adapt its policies and infrastructure to remain competitive amid rapid technological evolution. For generative AI and foundation models, measures should include mandating transparent disclosure, such as clear labeling and watermarking of AI-generated content as stipulated by the AI Act, and funding multilingual large language models that reflect Europe's diverse linguistic heritage. In the realm of quantum computing, integrating quantum accelerators with existing HPC resources and developing adaptive cyber-defense standards will be essential to harness its exponential potential while mitigating dual-use risks. Equally, advancing next-generation connectivity requires closing 5G deployment gaps, investing in 6G research, and establishing edge computing testbeds to support real-time AI applications. Finally, by harnessing synergies with complementary fields such as VR/AR, robotics, and biotechnology, Europe can unlock new markets and strengthen its strategic autonomy. Together, these measures will ensure that Europe's regulatory and technological frameworks remain agile and resilient in the face of emerging paradigms.

\section{Europe’s Path to AI Leadership - A Balanced Vision for the Future}

Europe stands at a pivotal crossroads. The global AI race is not merely a contest of technological prowess but a battle for values, sovereignty, and societal resilience. While the U.S. and China dominate through scale and speed, Europe’s unique opportunity lies in harmonizing its regulatory foresight, industrial ingenuity, and human-centric ethos into a cohesive strategy. This paper has outlined a roadmap to transform Europe from a reactive rule-setter into a proactive, innovation-driven leader capable of shaping the AI era on its own terms.

Europe’s strengths, its world-class research ecosystem, globally influential regulatory frameworks, and network of agile “hidden champions”, are formidable foundations. Yet these alone are insufficient. To thrive, Europe must urgently bridge critical gaps in semiconductor sovereignty, high-performance computing, and SME empowerment while fostering a culture of agile experimentation. The proposed European AI Coordination Council offers a blueprint for harmonizing standards without stifling innovation, marrying accountability with agility through sector-specific sandboxes and risk-calibrated audits. By prioritizing market-driven incentives over heavy-handed subsidies, Europe can unlock the latent potential of its SMEs, turning niche excellence into global leadership.

Equally vital is the cultivation of talent and public trust. Europe’s workforce must be equipped not only with technical skills but with the ethical literacy to navigate AI’s societal implications. Strategic immigration reforms, industry-aligned education, and lifelong reskilling programs will ensure a steady pipeline of innovators. Concurrently, transparent governance through AI labeling, citizen assemblies, and global partnerships will reinforce Europe’s role as a beacon of responsible innovation.

Critically, Europe must embrace multi-technology convergence as a strategic imperative. AI’s synergy with quantum computing, biotechnology, and next-gen connectivity will define future industries and geopolitical power. By embedding interoperability into its infrastructure and policies, Europe can lead in cross-cutting domains like green tech, precision medicine, additive manufacturing, and resilient supply chains.

The stakes could not be higher. Hesitation or fragmentation risks relegating Europe to a regulatory island that is respected but irrelevant in shaping the technologies reshaping our world. Conversely, decisive action anchored in Europe’s values, democratic accountability, privacy, and sustainability, can position it as the global standard-bearer for AI that empowers.

This vision demands more than ambition; it requires political courage, cross-border collaboration, and a willingness to rethink outdated paradigms. By uniting its regulatory clout with industrial agility, Europe can transcend the false dichotomy between innovation and ethics. The result will be a competitive, values-driven AI ecosystem that fuels economic growth, reinforces strategic autonomy, and exports a model of inclusive progress worldwide.

The time for incrementalism has passed. Europe must act with the urgency this moment demands, not merely to compete in the AI race, but to redefine its rules and strategy. By doing so, it will secure not just technological leadership but a future where innovation serves humanity, anchored in the principles that have long defined Europe’s identity.

\medskip
\noindent\textbf{Acknowledgements.} The authors acknowledge the insights from governmental bodies, industry experts, and academic institutions across Europe who have contributed to shaping this holistic AI governance vision.

\small
\balance
\bibliographystyle{unsrt}
\bibliography{references}

\end{document}